\documentstyle[12pt,amssymb,cite]{article}

\def\mprp{\mbox{\tiny $\bot$}}
\def\mprl{\mbox{\tiny $\|$}}

\title{
\begin{flushright}
{\normalsize Yaroslavl State University\\
             Preprint YARU-HE-01/05\\
             hep-ph/0108046} \\[5mm]
\end{flushright}
Photon-pair conversion into neutrinos \\
in a strong magnetic field}

\author{{A.V.~Kuznetsov and N.V.~Mikheev}\\ [7mm] 
{\small\it Division of Theoretical Physics, Department of Physics,}\\
{\small\it Yaroslavl State University, Sovietskaya 14,}\\
{\small\it 150000 Yaroslavl, Russian Federation}}

\date{}

\begin{document}

\maketitle

\begin{abstract}
A general analysis of the three-vertex loop amplitude in
a strong magnetic field, based on the asymptotic form of 
the electron propagator in the field, is performed.
In order to investigate the photon-neutrino process
$\gamma \gamma \to \nu \bar\nu $, the vertex combinations of the
scalar--vector--vector ($SVV$), pseudoscalar--vector--vector ($PVV$),
3--vector ($VVV$), and axial-vector--vector--vector ($AVV$)
types are considered. It is shown that only the $SVV$ amplitude grows
linearly with the magnetic field strength, while in the
other amplitudes, $PVV$, $VVV$, and $AVV$, the linearly growing
terms are cancelled.
The process $\gamma \gamma \to \nu \bar\nu$ is investigated 
in the left-right-symmetric extension of the standard model 
of electroweak interaction, where the effective scalar $\nu \nu e e$
coupling could exist. 
Possible astrophysical manifestations of the considered
process are discussed.
\end{abstract}
 
\vfill

\begin{center}
{\sl 
Submitted to Modern Physics Letters A}
\end{center}

\newpage

Strong magnetic fields which could be generated in the astrophysical 
cataclysms like a supernova explosion or a coalescence of neutron stars, 
make an active influence 
on quantum processes, thus allowing or enhancing the transitions 
which are forbidden or strongly suppressed in vacuum.
However, the magnetic field influences significantly the quantum processes 
only in the case when it is strong enough. 
There exists a natural scale for the field strength which is the so-called 
critical Schwinger value $B_e = m_e^2/e \simeq 4.41 \cdot 10^{13}$ G
(we use natural units in which $c = \hbar = 1$, hereafter 
$e$ is the elementary charge). 
Among others, the set of quantum processes is very interesting where only 
electrically neutral particles, such as neutrinos and photons, present 
in the initial and final states. The external field influence on these 
loop processes is provided by the sensitivity of the charged virtual fermion 
to the field. The electron plays the main role here as the particle with 
the maximal specific charge, $e / m_e$. 

In this letter we consider a three-vertex loop process
in a strong magnetic field, which is described by the Feynman 
diagram shown in Fig.1. Here the double lines correspond to the 
electron propagators constructed on the base of the exact solutions of the 
Dirac equation in a magnetic field. 

The intensively discussed process of this type is
the conversion of the photon pair into the neutrino - antineutrino pair, 
$\gamma \gamma\to \nu \bar\nu$, which is strongly suppressed in vacuum.
For this process, two vertices are vectors, e.g. $\Gamma_1 = \Gamma_2 = V$, 
and the third one can be vector and axial-vector, $\Gamma_3 = V,A$ 
(in the standard model), or it could be scalar and pseudoscalar, 
$\Gamma_3 = S,P$ (beyond the standard model). 
It is well-known (the so-called Gell-Mann 
theorem~\cite{Gell-M}), that for massless neutrinos, for both photons 
on-shell and in the local limit of the standard-model
weak interaction, the process
$\gamma \gamma\to \nu \bar\nu$ is forbidden.
With any deviation from the Gell-Mann theorem conditions, 
the process becomes allowed, e.g. for finite neutrino mass~\cite{Crew,Dode} 
or with the non-locality of the weak interaction taken into 
account~\cite{Dicus93}.

This process becomes also possible, in the frame of the standard model,
in the presence of external magnetic field. In the limit of relatively weak 
magnetic field, $B \ll B_e$, the process was considered in Refs.~\cite{Shai}.
Our aim is to investigate the process $\gamma \gamma\to \nu \bar\nu$ in the 
strong field limit, both in the standard model and beyond.
So, it is worthwhile to set $\Gamma_1$ and $\Gamma_2$ as 
vector vertices contracted with photons, keeping $\Gamma_3$ to be arbitrary.

The electron propagator in a magnetic field 
can be presented in the form 
\begin{eqnarray}
S(x,y) &=& e^{\mbox{\normalsize $i \Phi (x,y)$}}\, \hat S(x-y),
\label{Sxy}\\
\Phi(x,y) &=& - e \int \limits^y_x d\xi_\mu \left [ A_\mu(\xi) + \frac{1}{2} 
F_{\mu \nu}(\xi - y)_\nu \right ],
\label{Phi}
\end{eqnarray}

\noindent 
where $A_\mu$ is a 4-potential of the external uniform magnetic field. 
The translational invariant part $\hat S(x-y)$ of the propagator has several 
representations. We take its asymptotic form in the strong field limit, 
when the magnetic field strength is the maximal physical parameter:
\begin{eqnarray}
\hat S(X) &\simeq& 
\frac{i eB}{2 \pi} \exp (- \frac{eB X_{\mprp}^2}{4}) 
\int \frac{d^2 p}{(2 \pi)^2}\, \frac{(p\gamma)_{\mprl} + m}{p_{\mprl}^2 - m^2}
\Pi_{-}e^{-i (pX)_{\mprl}}, 
\label{Sa} \\
d^2 p &=& dp_0 dp_3, \quad \Pi_- = \frac{1}{2} (1 - i \gamma_1 \gamma_2),
\quad \Pi^2_- = \Pi_-, \quad [\Pi_-, (a \gamma)_{\mprl}] = 0.
\nonumber
\end{eqnarray}
 
\noindent 
The propagator was obtained in this form for the first time in 
Ref.~\cite{Skob}. 
Here $\gamma_\alpha$ are the Dirac matrices in the standard 
representation, the four-vectors with the indices $\bot$ and $\parallel$ 
belong to the Euclidean \{1, 2\} plane and the Minkowski \{0, 3\} plane 
correspondingly, 
when the field $\bf B$ is directed along the third axis. Then for arbitrary 
4-vectors $a_\mu$, $b_\mu$ one has
\begin{eqnarray}
a_{\mprp} &=& (0, a_1, a_2, 0), \quad  a_{\mprl} = (a_0, 0, 0, a_3), 
\nonumber \\
(ab)_{\mprp} &=& (a \Lambda b) =  a_1 b_1 + a_2 b_2 , \quad 
(ab)_{\mprl} = (a \widetilde \Lambda b) = a_0 b_0 - a_3 b_3, 
\label{ab}
\end{eqnarray}

\noindent where the matrices are introduced
$\Lambda_{\alpha \beta} = (\varphi \varphi)_{\alpha \beta}$,\,  
$\widetilde \Lambda_{\alpha \beta} = 
(\tilde \varphi \tilde \varphi)_{\alpha \beta}$, connected by the relation 
$\widetilde \Lambda_{\alpha \beta} - \Lambda_{\alpha \beta} = 
g_{\alpha \beta} = diag(1, -1, -1, -1)$, 
$\varphi_{\alpha \beta} =  F_{\alpha \beta} /B$ 
is the dimensionless tensor of the external 
magnetic field, 
${\tilde \varphi}_{\alpha \beta} = \frac{1}{2} \varepsilon_{\alpha \beta
\mu \nu} \varphi_{\mu \nu}$ 
is the dual tensor, 
the indices of four-vectors and tensors standing inside the 
parentheses are contracted consecutively, e.g. 
$(a \Lambda b) = a_\alpha \Lambda_{\alpha \beta} b_\beta$.

In spite of the translational and gauge noninvariance of the phase 
$\Phi(x, y)$ in the propagator~(\ref{Sxy}), the total phase of three 
propagators in the loop is translational and gauge invariant
\begin{eqnarray} 
\Phi(x, y) + \Phi(y, z) + \Phi(z, x) = - \frac{e}{2} (z - x)_\mu 
F_{\mu \nu} (x - y)_\nu. \nonumber
\end{eqnarray}

A general amplitude of the process described by Fig.1 takes the form
\begin{eqnarray}
{\cal M} &=& 
e^2 g_3 \int d^4 X\, d^4 Y\, Sp \{(j_3 \Gamma_3)
\hat S(Y) (\varepsilon_2 \gamma)
\hat S(-X-Y) (\varepsilon_1 \gamma) \hat S(X)\} \times 
\nonumber \\
&\times& e^{- i e\,(X F Y)/2}\; e^{i (k_1 X - k_2 Y)} + 
(\gamma_1 \leftrightarrow \gamma_2),
\label{M}
\end{eqnarray}

\noindent 
where 
$X = z - x, \, Y = x - y$, 
$\Gamma_3$ is the matrix corresponding to 
the abitrary ($S,P,V$ or $A$) vertex, 
$g_3$ is the coupling constant, 
$j_3$ is the neutrino current in the momentum space, 
$\varepsilon_1,\; k_1$ and $\varepsilon_2,\;k_2$ are the polarization 
vectors and the 4-momenta of initial photons.

Substituting the propagator~(\ref{Sxy}),~(\ref{Phi}), and~(\ref{Sa}) into the 
amplitude one obtains that two parts of it which differ by the photon 
interchange, are proportional to the field strength $B$
\begin{eqnarray}
{\cal M} &\simeq& 
- \frac{i \,\alpha \,g_3 \,e B}{(4 \pi)^2} 
\exp \left\{- \frac{\vec k_{1\mprp}^2 + \vec k_{2\mprp}^2 
+ (\vec k_1 \vec k_2)_{\mprp}}{2eB}\right\} \; 
\exp \left\{- i \frac{(k_1 \varphi k_2)}{2eB}\right\} \times
\nonumber \\
& \times& \int d^2 p \, Sp \{(j_3 \Gamma_3)
S_{\mprl}(p+k_2) (\varepsilon_2 \gamma)
S_{\mprl}(p) (\varepsilon_1 \gamma) S_{\mprl}(p-k_1)\} +
\nonumber \\
&+& 
(\gamma_1 \leftrightarrow \gamma_2),
\label{M2}
\end{eqnarray}

\noindent 
where $S_{\mprl}(p) = 2 \Pi_{-} ((p\gamma)_{\mprl} + m)/(p_{\mprl}^2 - m^2)$.
It should be noted, that in the amplitude~(\ref{M2}) 
the projecting operators $\Pi_-$ separate out  
the photons of only one polarization $(\perp)$ of the two
possible (in Adler's notation~\cite{Ad71})
$$
\varepsilon^{(\mprl)}_{\alpha} \sim F_{\alpha\beta} k_{\beta},
\quad 
\varepsilon^{(\mprp)}_{\alpha} \sim \tilde F_{\alpha\beta} k_{\beta}.$$

Using the standard procedure one can transform the trace in the second 
term of Eq.~(\ref{M2}) with the photon interchanged into the trace of the 
first term, however, with change of sign for $\Gamma_3 = P,V,A$ 
(and the factor $\sin[(k_1 \varphi k_2)/2eB]$ arises in the resulting 
amplitude) and without change of sign for $\Gamma_3 = S$ 
(and the factor $\cos[(k_1 \varphi k_2)/2eB]$ appears after summation).
So, when the magnetic field strength is the maximal physical parameter, 
$e B \gg \vec k_{\mprp}^2, k_{\mprl}^2$, only the amplitude with 
the scalar vertex grows linearly with the field. 

The effective scalar $\nu \nu e e$ interaction arises in the standard 
model extensions with broken left-right (LR) symmetry~\cite{Lipm67} 
with mixing of the bosons coupled with left and 
right charged weak currents~\cite{Beg77}. The $\nu e W$ interaction in the 
model has the form
\begin{eqnarray}
{\cal L} &=& \frac{g}{2\sqrt{2}}\;\bigg\lbrace
\left[\bar e \gamma_\alpha
\left( 1 - \gamma_5 \right) \nu \right] 
\left(W_1^\alpha \cos \zeta + W_2^\alpha \sin \zeta \right)
+
\nonumber \\
&+&
\left[\bar e \gamma_\alpha
\left( 1 + \gamma_5 \right) \nu \right] 
\left(- W_1^\alpha \sin \zeta + W_2^\alpha \cos \zeta \right)
+ h.c.
\bigg\rbrace,
\label{L_LR}
\end{eqnarray}

\noindent 
where $W_{1,2}$ are the mass eigenstates, $\zeta$ is the mixing angle.
Existing indirect limits on the LR model parameters extracted from 
low-energy accelerator experiments is~\cite{RPP_00}
\begin{eqnarray}
M_{W_2} > 715 \; GeV, \quad \zeta < 0.013.
\label{LR_limit1}
\end{eqnarray}

\noindent
Due to the smallness of the mixing, $W_2$ is almost $W_R$.
Indirect limit on the parameters from astrophysical data (SN1987A) 
is even more stringent.
In combination with the accelerator data it gives~\cite{Barbi}
\begin{eqnarray}
M_{W_R} > 23 \; TeV, \quad \zeta < 10^{-5}.
\label{LR_limit2}
\end{eqnarray}

\noindent
Taking into account the smallness of the mixing angle and of the 
mass ratio $M_{W_L} / M_{W_R}$, for the effective scalar part of 
the $\nu \nu e e$ interaction one obtains
\begin{eqnarray}
{\cal L}_{eff}^{(s)} \simeq - 4 \; \zeta \; \frac{G_F}{\sqrt{2}} 
\left(\bar e e \right)
\left(\bar \nu \nu \right).
\label{L_eff}
\end{eqnarray}

\noindent 
With the Lagrangian~(\ref{L_eff}), there exist two channels of the conversion 
of the photon pair into the neutrino pair, namely,
\begin{eqnarray}
\gamma \gamma\to (\nu_e)_L (\bar\nu_e)_L, \quad
\gamma \gamma\to (\nu_e)_R (\bar\nu_e)_R. 
\label{proc}
\end{eqnarray}

\noindent 
Here $(\nu_e)_R$ and $(\bar\nu_e)_L$ are the ``sterile'' states with respect
to the standard weak interaction, which can escape from the hot and dense 
stellar interior. 

Substitution of $\Gamma_3 = 1$, $g_3 = - 4 \; \zeta \; G_F/\sqrt{2}$ and 
$j_3 = [\bar \nu(p_1) \nu(-p_2)]$, resulting from Eq.~(\ref{L_eff}), into  
the amplitude~(\ref{M2}) and integration give in the strong field limit
\begin{eqnarray}
{\cal M} &=& - \frac{4 \alpha}{\pi}\;\frac{B}{B_e}\;\frac{\zeta\,G_F}{\sqrt{2}\,m_e}
\;[\bar \nu(p_1) \nu(-p_2)]\;
\bigg\lbrace \left(f_1^{(\mprp)} f_2^{(\mprp)}\right)
{\cal F}_1 \left( \frac{q_{\mprl}^2}{m_e^2}, \frac{k_{1\mprl}^2}{m_e^2}, 
\frac{k_{2\mprl}^2}{m_e^2}\right) +
\nonumber \\
&+& 4 \; \frac{(k_{1\mprl} f_1^{(\mprp)} f_2^{(\mprp)} k_{2\mprl})}{q_{\mprl}^2} \;
{\cal F}_2 \left( \frac{q_{\mprl}^2}{m_e^2}, \frac{k_{1\mprl}^2}{m_e^2}, 
\frac{k_{2\mprl}^2}{m_e^2}\right) \bigg\rbrace, 
\label{M3}
\end{eqnarray}

\noindent 
here $q_{\mprl} = k_{1\mprl} + k_{2\mprl}$. Remind, that 
$k_{1,2\mprl} = (\omega_{1,2},\;0,\;0,\;k_{1,2 z})$. 
$f_{\alpha\beta}$ are the photon field tensors
$$f^{(\mprp)}_{\alpha\beta} = k_{\alpha\mprl} \varepsilon^{(\mprp)}_{\beta} - 
k_{\beta\mprl} \varepsilon^{(\mprp)}_{\alpha}.$$
The functions introduced in Eq.~(\ref{M3}) are the following
\begin{eqnarray}
&& {\cal F}_1 (z,s,t) = \int\limits_0^1 x d x \int\limits_0^1 d y 
\frac{1 - 4 x^2 y (1 - y)}{a^2},
\nonumber \\
&& {\cal F}_2 (z,s,t) = \int\limits_0^1 x (1 - x) (1 - 2 x) d x 
\int\limits_0^1 \frac{d y}{a^2},
\nonumber \\
&& a = 1 - z x^2 y (1 - y) - s x (1 - x) y - t x (1 - x) (1 - y).
\label{fun}
\end{eqnarray}

\noindent 
The amplitude~(\ref{M3}) is manifestly gauge-invariant. 

The cross-sections for both processes~(\ref{proc}) are equal,
$\sigma_{LL} = \sigma_{RR} \equiv \sigma$. It takes a simple form 
in the two limiting cases:

\begin{itemize}
\item[i)] 
low photon energies, $\omega \lesssim m_e$
\begin{eqnarray}
\sigma \simeq \frac{2\, \alpha^2\, G_F^2\, \zeta^2}{9 \pi^3}
\; \left(\frac{B}{B_e}\right)^2
\frac{k_{1\mprl}^2 k_{2\mprl}^2}{m_e^2},
\label{sig1}
\end{eqnarray}

\item[ii)] 
high photon energies, $\omega \gg m_e$, in the leading log approximation:
\begin{eqnarray}
\sigma \simeq \frac{2\, \alpha^2\, G_F^2\, \zeta^2}{\pi^3}
\; \left(\frac{B}{B_e}\right)^2
\frac{m_e^6}{k_{1\mprl}^2 k_{2\mprl}^2} \;
\ln^2 \frac{k_{1\mprl}^2 k_{2\mprl}^2}{m_e^4}.
\label{sig2}
\end{eqnarray}
\end{itemize}

The observable value in astrophysics is the stellar energy-loss 
from unit volume in unit time
due to the neutrino escape (neutrino emissivity). In our case it 
can be written in the form
\begin{eqnarray}
Q = {1 \over 2} 
\; \int \frac{d^3 k_1}{(2 \pi)^3} \; \frac{1}{e^{\omega_1/T} - 1}
\; \int \frac{d^3 k_2}{(2 \pi)^3} \; \frac{1}{e^{\omega_2/T} - 1}
\; (\omega_1 + \omega_2) \; \frac{(k_1 k_2)}{\omega_1 \omega_2} 
\; \sigma. 
\label{Qdef} 
\end{eqnarray}

\noindent
Here $T$ is the temperature of the photon gas.
It is taken into account, that photons of only one polarization 
participate in the process, and only one (anti)neutrino from the pair can 
escape while the other one is trapped by the hot and dense 
stellar interior. 

In the low temperature case, $T \lesssim m_e$, 
substituting~(\ref{sig1}) into~(\ref{Qdef}), one obtains
\begin{eqnarray}
Q_{(B)} \simeq 2.5 \cdot 10^{13} \; \frac{\mbox{erg}}{\mbox{cm}^3 \; \mbox{s}} 
\; \left(\frac{\zeta}{0.013}\right)^2
\left(\frac{B}{B_e}\right)^2
\left(\frac{T}{m_e}\right)^{11}.
\label{Q2} 
\end{eqnarray}

\noindent
This should be compared with the contributions into the neutrino 
emissivity of the process $\gamma \gamma\to \nu \bar\nu$ caused 
by another mechanisms. 
For instance, the emissivity 
due to the finite neutrino mass is~\cite{Dode}
\begin{eqnarray}
Q_{(m_\nu)} \simeq 0.4 \cdot 10^5 \; \frac{\mbox{erg}}{\mbox{cm}^3 \; \mbox{s}} 
\; \left(\frac{m_\nu}{1 \,\mbox{eV}}\right)^2
\left(\frac{T}{m_e}\right)^{11}.
\label{Q3} 
\end{eqnarray}

\noindent
On the other hand, in the case of non-locality of the weak interaction, 
investigated in Ref.~\cite{Dicus93}, one can estimate the emissivity, 
which is suppressed by the factor $(m_e/M_W)^4$: 
\begin{eqnarray}
Q_{(NL)} \; \sim  
\; 10 \; \frac{\mbox{erg}}{\mbox{cm}^3 \; \mbox{s}} 
\; \left(\frac{T}{m_e}\right)^{13}.
\label{Q4} 
\end{eqnarray}

\noindent
It is seen that for $B \gtrsim B_e$, and for mixing at the level of 
$\zeta \sim 10^{-5}$, the field-induced mechanism 
of the reaction $\gamma \gamma\to \nu \bar\nu$ strongly dominates 
all the other mechanisms.

In the high temperature case, $T \gg m_e$, substituting~(\ref{sig2}) 
into~(\ref{Qdef}), one obtains
\begin{eqnarray}
Q_{(B)} \simeq 0.4 \cdot 10^{12} \; \frac{\mbox{erg}}{\mbox{cm}^3 \; \mbox{s}} 
\; \left(\frac{\zeta}{0.013}\right)^2
\left(\frac{B}{B_e}\right)^2
\left(\frac{T}{m_e}\right)^3
\ln^5 \frac{T}{m_e}.
\label{Q5} 
\end{eqnarray}

In order to make a numerical estimation, let us consider the Supernova 
explosion with generation of very strong magnetic field $B \sim 10^3\; B_e$, 
see e.g.~\cite{Bis70}, with the temperature $T \sim 35\;MeV$ which 
is believed to be typical for the Supernova core~\cite{Raff}, and 
$V \sim 10^{18}\;\mbox{cm}^3$. For the contribution of the considered 
field-enhanced process $\gamma \gamma\to \nu \bar\nu$ into the neutrino 
luminosity we obtain
\begin{eqnarray}
\frac{dE}{dt} \sim  
2 \cdot 10^{44} \; \frac{\mbox{erg}}{\mbox{sec}} 
\; \left(\frac{\zeta}{0.013}\right)^2.
\label{lum} 
\end{eqnarray}

\noindent
It is too small if compared with the typical Supernova neutrino luminosity
$10^{52} \; \mbox{erg}/\mbox{sec}$.

In summary,
we have performed a general analysis of the three-vertex 
loop in a strong magnetic field and have 
considered the photon-neutrino process 
$\gamma \gamma\to \nu \bar\nu$. 
It is shown that various types of the neutrino -
electron effective interactions lead to different dependences of the 
amplitudes on the field strength.
The effective scalar $\nu\nu e e$ coupling which exists in the extensions 
of the standard model with broken left-right symmetry, leads to 
the enhancement of the process $\gamma \gamma \to \nu \bar\nu$ by strong 
external magnetic field.
For the field strength $B \gtrsim B_e$ this mechanism could dominate 
other mechanisms of the process $\gamma \gamma\to \nu \bar\nu$ 
in the neutrino emissivity. 
However, its contribution into the Supernova energetics is rather small.

\medskip

This work was supported in part by the Russian Foundation for Basic 
Research under the Grant N~01-02-17334
and by the Ministry of Education of Russian Federation under the 
Grant No. E00-11.0-5.



\begin{minipage}[t]{100mm}

\unitlength=1.00mm
\special{em:linewidth 0.4pt}
\linethickness{0.4pt}

\vspace*{35mm}

\begin{picture}(100.00,17.00)(-20,10)

\put(30.00,38.00){\circle{16.00}}
\put(30.00,38.00){\circle{13.00}}
\put(36.6,38.00){\circle*{1.0}}
\put(24.30,34.30){\circle*{1.0}}
\put(24.30,41.70){\circle*{1.0}}

\put(24.00,29.00){\makebox(0,0)[cc]{$\Gamma_1$}}
\put(24.00,47.00){\makebox(0,0)[cc]{$\Gamma_2$}}
\put(39.50,41.00){\makebox(0,0)[cc]{$\Gamma_3$}}

\put(37.0,18.00){\makebox(0,0)[cc]{Fig. 1}}

\multiput(24.30,34.30)(-4.20,-4.20){3}{\line(-1,-1){3.6}}
\multiput(24.30,41.70)(-4.20,4.20){3}{\line(-1,1){3.6}}
\multiput(37.0,38.00)(4.20,0){3}{\line(1,0){3.6}}

\end{picture}

\end{minipage}

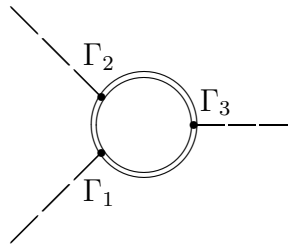
\end{document}